\documentclass{optica-article}

\journal{opticajournal} % for journals or Optica Open

\articletype{Research Article}
\usepackage{pdfpages}
\usepackage{lineno}
\begin{document}

\title{Poisson wavefront imaging in photon-starved scenarios}

\author{Seungman Choi,\authormark{1,†} Peter Menart,\authormark{1,†} Andrew Schramka,\authormark{1} Leif Bauer,\authormark{1} Vaneet Aggarwal\authormark{1} In-Yong Park \authormark{2} and Zubin Jacob \authormark{1,*}}

\address{
\authormark{1}Elmore Family School of Electrical and Computer Engineering, Birck Nanotechnology Center, Purdue University, West Lafayette, Indiana 47907, USA\\
\authormark{2}Emerging Research Instruments Group, Strategic Technology Research Institute, Korea Research Institute of Standards and Science (KRISS), 267 Gajeong-ro, Yuseong-gu, Daejeon 34113, South Korea.\\
\authormark{†}
These two authors contributed equally to this work.\\
}

\email{\authormark{*}zjacob@purdue.edu} %% email address is required; see note below about the corresponding author designation

% use {asbstract*} to suppress the copyright line. Copyright information will be added in production

\begin{abstract*} 
Low-photon phase imaging is essential in applications where the signal is limited by short exposure times, faint targets, or the need to protect delicate samples. We address this challenge with Poisson Wavefront Imaging (PWI), an optimization-based method that incorporates Poisson photon statistics and a smoothness prior to improve wavefront reconstruction. By using multiple spatial light modulator's phase patterns, PWI enhances Fisher information, boosting theoretical accuracy and regularizing the retrieval process effectively. 
In simulations, PWI approaches the theoretical phase error limit, and in experiments it reduces phase error by up to 1.6$\times$ compared to the Gerchberg-Saxton algorithm, achieving 1.8$\times$ higher resolution wavefront imaging in low photon regime. This method advances photon-limited imaging with applications in astronomy, semiconductor metrology, and biological systems.

\end{abstract*}

%%%%%%%%%%%%%%%%%%%%%%%%%%  body  %%%%%%%%%%%%%%%%%%%%%%%%%%
\section{Introduction}
 Low-photon wavefront sensing is essential for correcting aberrations \cite{GuyonExAO, davies2012adaptive, choi2025photon} and revealing structural details \cite{QuantPhaseImg, sun2017resolution, pan2020high, won2025programmable} in various  imaging applications such as astronomy, semiconductor metrology, and biology. In these fields, multiple factors restrict the available signal. High illumination can damage sensitive biological samples \cite{magidson2013circumventing, kilian2018assessing}. Coherent EUV to X-ray imaging suffer  from low signal due to the nonlinear light generation processes \cite{kim2022compensation, kim2023coherent}. In astronomical imaging, distant or small objects such as stars or orbiting  debris appear faint \cite{zheng2015improved, sun2015algorithms, choi2024telescope}. Additionally, short exposure times are often required for dynamic biological imaging \cite{daetwyler2016fast, winter2014faster} or for observing through dynamic atmospheric turbulence \cite{mao2020image, feng2023neuws, bao2021quantum}. These challenges demand advanced wavefront sensing that maintains high performance in extreme photon-limited conditions.

Two main strategies address this problem: direct methods with specialized sensors and indirect inference via optimization.
Among direct methods, the Shack-Hartmann wavefront sensor (SHWFS) is widely used in adaptive optics for real-time corrections of dynamic aberrations. 
By tracking focal spot displacements in a lenslet array, it quickly measures wavefront gradients, enabling real-time measurement \cite{thomas2006comparison}. However, shot noise degrades SHWFS performance in low-light conditions, requiring optimization-based estimation to handle low-photon statistics \cite{li2018centroid}, which diminishes its real-time advantage. Its spatial resolution is also limited by the required microlens array, typically less than 100$\times$100 samples, making it less suitable for strong aberrations \cite{saita2015holographic}. 

Indirect wavefront imaging, or phase retrieval, treats wavefront estimation as an inverse problem: $I_i = |A_i x|^2$, where $i=\{1,\cdots, M\}$ indexes the measurement, $x$ is the complex electric field to be estimated (with $n$ pixels), $I_i$ is the intensity measurement (with $m$ pixels), and $A_i$ is the $m\times n$ measurement matrix describing the optical system and light  propagation. A unique solution for $x$ typically requires  $mM\gtrsim4n$  when $A$ is full rank \cite{CompImgML}. The reconstruction is formulated as: 
\begin{equation} 
x = \operatorname*{argmin}_{x \in \mathbb{C}^n} \left[ \sum_{i=1}^{M} F(I_i, u_i) + R(x) \right] \quad \text{s.t.}  \quad C_i(x, u_i)=0, \label{eq:IntroEq1} 
\end{equation} 
The fidelity term $F(I_i, u_i)$ enforces consistency between the measurements $I_i$ and the predicted detector fields $u_i$. The regularization term $R(x)$ incorporates prior knowledge of the target field $x$, while the constraint $C_i(x, u_i)$ enforces physical consistency, such as wave propagation. 
Conventional methods like Gerchberg-Saxton \cite{seifert2023maximum} use a mean square error fidelity term that matches Gaussian noise statistics but is inconsistent with Poisson shot noise in low-light scenarios. Multiple measurements $\{I_i\}_{i=1, \dots, M}$ can improve reconstruction accuracy \cite{candes2015phase, gross2017improved, wu2019wish} or extend the field of view, as in ptychography \cite{kim2023coherent, seifert2023maximum}.
Further improvements have been achieved through spectral initialization \cite{maillard2022construction}, denoising methods such as wavelet \cite{qin2024complex} and BM3D \cite{katkovnik2017computational}, and priors that impose spatial constraints or sparsity \cite{ibanez2024retrieval, hyland2015noise, zhou2020low}. Despite these  advances, most low-photon phase retrieval techniques
have only been demonstrated in simulations.

Recently, deep learning-based methods have shown superior performance at low photon levels using neural networks and training data as strong regularizers \cite{BarbLowPhot, sinha2017lensless}. However, their reliance on representative training data poses challenges when such data is either unavailable or poorly aligned with target conditions. 

In this paper, we propose Poisson Wavefront Imaging (PWI) for low-photon wavefront imaging. PWI employs a spatial light modulator (SLM) to implement multiple phase diversity patterns, which act as strong, user-defined regularizers $C_i(x,u_i)=0$ in equation (\ref{eq:IntroEq1}). This approach enables high-quality reconstruction without reliance on data-driven priors. 
We prove that the applied phase patterns enhance Fisher information, improving sensitivity to small phase variations and achieving higher theoretical accuracy than the SHWFS under the same photon budget.
Poisson photon statistics are incorporated into the fidelity term $F(I_i, u_i)$, while total variation (TV) priors $R(x)$ improve image smoothness \cite{li2022poisson, katkovnik2017computational}. Simulations and experiments confirm that PWI offers a robust solution for low-photon wavefront sensing. 

\section{Methods}
The PWI framework comprises two main components: a coded-detection architecture that provides phase diversity, and an optimization algorithm that reconstructs the complex field using a Poisson likelihood with TV regularization in the alternating direction method of multipliers (ADMM) approach. We summarize both elements below.

\begin{figure}[t]
\centering\includegraphics[width=0.60\textwidth]{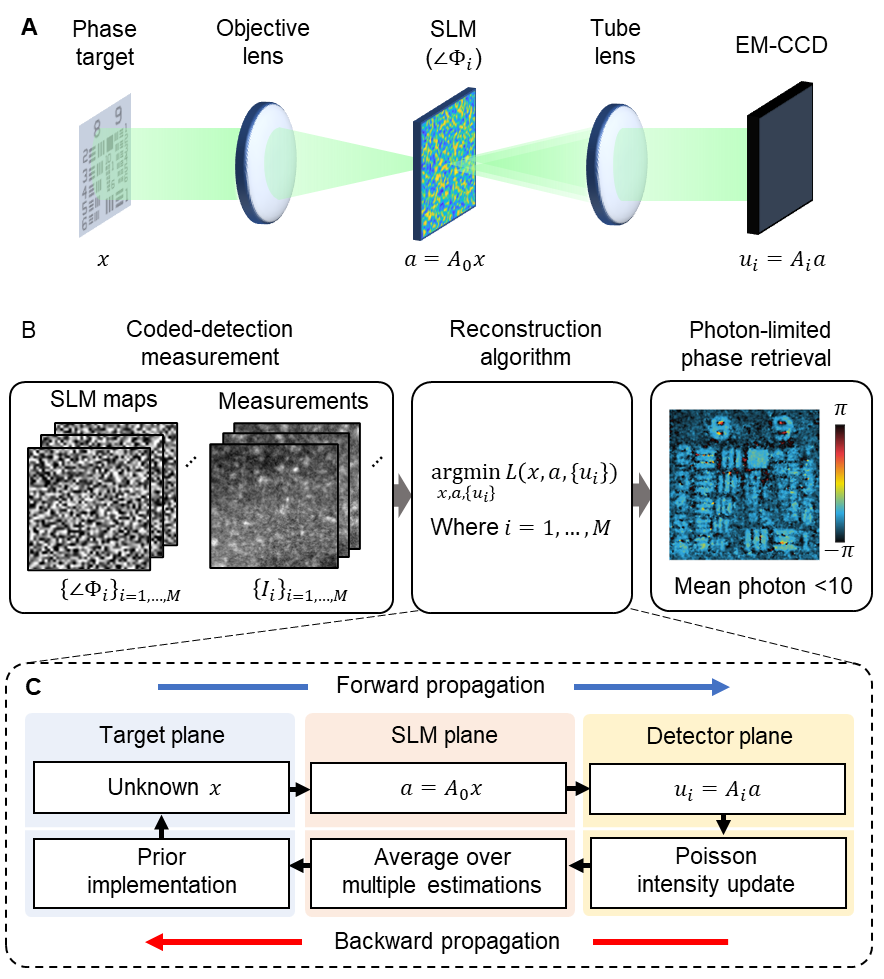}
\caption{Overview of PWI frameworks. A. Simplified diagram of coded-detection setup. B. PWI using multiple phase diversity patterns $\{\Phi_i\}_{i=1, \dots, M}$ and corresponding measured images $\{I_i\}_{i=1, \dots, M}$. C. Iterative reconstruction of the complex e-fields through forward and backward propagation.}
\label{Fig1}
\end{figure}

\subsection{PWI hardware architecture}
Figure \ref{Fig1}A illustrates the PWI setup with the complex field at three key planes: the target plane $x$, the SLM plane $a$, and the detector planes $\{u_i\}_{i=1,\dots,M}$ corresponding to different SLM phase patterns $\{\Phi_i\}_{i=1,\dots,M}$.  
These fields are related by the angular spectrum method (ASM):

\begin{equation}
a = A_0x, \quad u_i = A_ia
\label{eq2}
\end{equation} 
where $A_0$ models wave propagation from the phase target to the SLM, and $A_i$ includes both propagation from the SLM to the detector and the phase modulation induced by the $i$-th SLM pattern $\Phi_i$.  
The corresponding intensity measurements $\{I_i\}_{i=1,\dots,M}$ are recorded at the detector as $I_i = |u_i|^2$.  
This coded-detection forward model directly describes the PWI measurement process and provides the phase diversity that enables robust phase retrieval in the low-photon regime (Figure \ref{Fig1}B).

\subsection{PWI reconstruction algorithm}
Given this forward model, PWI recovers the complex target field by solving a Poisson-likelihood inverse problem with the TV prior. Let $u_{ik}$ denote the complex field at detector pixel $k$ for the $i$-th measurement, and $I_{ik}$ the corresponding photon counts. We formulate the reconstruction as:

\begin{equation}
\begin{split}
&\operatorname*{argmin}_{\{u, a, x, T_x\}} \sum^{K}_{k=1} \sum^{M}_{i=1} \{- 2I_{ik}\log(|u_{ik}|) + |u_{ik}|^2 \} +\tau ||T_x||_1 \\
& \quad s.t. \quad T_x = \Psi x, \quad u_i = A_ia, \quad a = A_0 x.
\end{split}
\label{eq3}
\end{equation}
Here, the main fidelity term in the braces models the Poisson statistics of the measurements, $\Psi$ denotes the TV operator, $T_x$ is the corresponding auxiliary variable, and $\tau$ controls the strength of the TV regularization.  
The objective combines a Poisson negative log-likelihood fidelity term with an $\ell_1$ TV penalty, while the constraints enforce physically accurate wave propagation between the target, SLM, and detector planes.  
In this setting, TV regularization acts as an empirical prior that preserves sharp phase features while suppressing noise under severe photon limitations, making it particularly effective for our low-SNR imaging problem~\cite{strong2003edge,shen2018improved}.

We solve this constrained optimization problem using the ADMM. By introducing auxiliary variables and Lagrange multipliers, ADMM decomposes equation \eqref{eq3} into a sequence of simpler sub-problems with closed-form updates for $u$, $a$, $x$, $T_x$, and the dual variables. The resulting update rules are summarized in Table~1 and follow the ADMM-based alternating update scheme \cite{liang2017phase,katkovnik2017phase}. Additional derivation details are provided in the Supplementary Material.

\begin{table}%[b]
\renewcommand{\arraystretch}{1.3}
\caption{Alternating update rule for back propagation}
\label{tab:ADMMUpdate}
\begin{footnotesize} % Smaller font size for the entire table
\begin{tabular}{lc}
\hline
\textrm{Plane} & \textrm{Update rule} \\
\hline
Detector & 
\begin{tabular}[c]{@{}c@{}} 
$u_{i} = |u_i|e^{j\angle (g_i)}$, where $g_i= A_ia+\lambda_{1i}$ \\
$|u_i| = \left(4 + 2\gamma_1 \right)^{-1} \left( \gamma_1|g_i| + \sqrt{\gamma_1^2|g_i|^2 + 8I_i(2+\gamma_1)} \right)$ 
\end{tabular} \\
\hline
Pupil & 
$a = \left( \gamma_1 M + \gamma_3 \right)^{-1} \left( \gamma_1 \sum^M_{i=1} A_i^H(u_i-\lambda_{1i}) + \gamma_3 (A_0 x-\lambda_3) \right)$ \\
\hline
Target &
\begin{tabular}[c]{@{}c@{}} 
$x = \left( \gamma_2 \Psi^H \Psi + \gamma_3 \right)^{-1} \left( \gamma_2 \Psi^H (T_x-\lambda_2) + \gamma_3 A_0^H (a+\lambda_3) \right)$ \\
$T_x = \max(0, |\Psi x| - {\tau}/{\gamma_2})e^{j \angle(\Psi x)}$
\end{tabular} \\
\hline
Dual variables (Lagrange multipliers) &
\begin{tabular}[c]{@{}c@{}} 
$\lambda_{1i} = \lambda_{1i} +\gamma_1(A_ia - u_i)$ \\
$\lambda_{2} = \lambda_{2} +\gamma_2(\Psi x - T_x)$ \\
$\lambda_{3} = \lambda_{3} +\gamma_3(a - A_0x)$ \\
\end{tabular} \\
\hline
\multicolumn{2}{l}{
\parbox{\linewidth}{
\footnotesize{
\vspace{0.2cm}
$\gamma_1, \gamma_2, \gamma_3$: Penalty coefficients,
$\Psi$: TV operator, 
$\tau$: Controls smoothness,
$T_x$: Auxiliary variable for TV regularization, 
$\angle(q)$: Phase of \(q\).
}
}
} \\
\hline
\end{tabular}
\end{footnotesize}
\end{table}

Figure \ref{Fig1}C summarizes each PWI iteration as a deterministic forward-backward propagation cycle. 
First, a forward propagation step updates the detector-plane fields $\{u_i\}$ from the current target estimate $x$ via the propagation operators in equation \eqref{eq2}.  
Then, a backward propagation step refines the variables by applying the update rule listed in Table \ref{tab:ADMMUpdate}. 

The computational bottleneck of our solver is the ASM operator, which performs two FFTs per propagation. Consequently, the cost per PWI iteration is $O(MK\log K)$. In contrast, stochastic gradient methods include additional overhead to track gradient statistics---particularly when using momentum-based optimizers such as Adam---resulting in higher per-iteration cost.
The proposed PWI solver achieves a 1.7$\times$ speed improvement over Adam optimizer along with substantially more accurate amplitude and phase reconstruction. A detailed convergence comparison is provided in the Supplementary Material.

\section{Results and discussion}
To assess the suitability of the PWI setup for low-photon phase imaging, we first compare the achievable theoretical accuracy limit of PWI and SHWFS, a widely used wavefront sensing method in ground-based optical observation (Figure \ref{Fig2}). Unlike PWI, which provides phase information directly at the detector pixel resolution, SHWFS estimates phase from lenslet-induced PSF centroid shifts and therefore requires much denser detector sampling to achieve the same spatial resolution. Accordingly, in our model, the SHWFS detector is assumed to have 11×11 times higher pixel density (Figure \ref{Fig2}A). 
We evaluate the accuracy limits of each method using the signle photon Fisher information matrix (FIM) and Cramer-Rao lower bound (CRLB). The parameter of interest is the phase at each pixel, denoted as $\Theta = \{\theta_p\}_{p=1,\dots, P}$, where $P=1600$ for a 40 $\times$ 40 phase image. The simulated phase forms the letter ``P'' with uniform amplitude. 
The FIM $J$ quantifies the information content of each system:
\begin{equation}
J_{pq} = \sum_{k=1}^K \sum_{i=1}^M \frac{1}{\mu_{ik}} \frac{\partial \mu_{ik}}{\partial \theta_p} \frac{\partial \mu_{ik}}{\partial \theta_q},
\label{eq:fisher_matrix}
\end{equation}
Where $K$ is the total number of detector pixels, $M=10$ is the number of coded-detection measurements, and $\mu_{ik}$ is the mean single-photon intensity at detector pixel $k$ for the $i$-th measurement. Since the measurements are invariant to global phase shift (piston), the FIM is rank-deficient.
To account for this, we fix the phase in the last pixel $p=P$, and remove the corresponding row and column from $J$, and invert the reduced matrix to obtain the CRLB \cite{bandeira2014saving}: $\sigma^2_p=[J^{-1}]_{pp}$, where $p=\{1,\cdots, P-1\}$. 
These diagonal elements of CRLB matrix give the variance for each phase element, and the mean standard deviation is: 
\begin{equation}
\sigma (N) =\sqrt{\frac{1}{N(P-1)}\sum_p \sigma_{p}^2}
\end{equation}
where $N$ is the photon count.
This represents the theoretical phase retrieval accuracy at photon level $N$, assuming an unbiased estimator.

\begin{figure} %[ht!] h: here, t: top, !: priority
\centering\includegraphics[width=0.60\textwidth]{}

\caption{
Theoretical accuracy vs. reconstruction performance for SHWFS and PWI.
A. Simulation setups for PWI and SHWFS, illustrating their detection architectures. 
B. Dashed curves show the square root of the Cramer–Rao lower bound, representing the theoretical phase precision for the SHWFS (blue) and for PWI using either random phase diversity (red) or a flat phase pattern (black, serving as the no-diversity reference). Solid curves show the corresponding reconstruction phase RMSE: SHWFS (blue) and PWI with (orange) and without (red) TV regularization. The PWI reconstruction curves (solid orange and red) should both be compared to the dashed red CRLB. Note that due to phase wrapping, the maximum attainable RMSE is constrained to approximately 0.28 (green dash–dot line).
C. Reconstructed phase images at 10,000, 158, and 1.58 mean photons per pixel for PWI with TV and SHWFS; bottom-row insets show SHWFS intensity measurements.
}
\label{Fig2}
\end{figure}
Figure \ref{Fig2}B shows the CRLB analysis for each setup, with dashed lines indicating the theoretical minimum uncertainty $\sigma$. To account for phase wrapping, we adjust the uncertainty bound following the approach in \cite{huang2010mse}. As photon counts decrease, the uncertainty bounds rise sharply, illustrating the fundamental challenge of low-photon wavefront imaging.
Among the tested methods, PWI with random phase patterns achieves a substantially lower uncertainty bound than both the flat phase pattern and the SHWFS, indicating higher sensitivity to phase changes.
%This improvement arises in part because, in the random phase setup, a phase change at a single target pixel influences a larger portion of the intensity measurement compared to the other configurations.
The solid lines in Figure \ref{Fig2}B present the reconstructed phase RMSE for the SHWFS and the random-phase PWI with and without the TV prior.
The SHWFS performs worst, with RMSE close to its theoretical bound.
PWI without the TV prior performs better but remains above its CRLB limit.
PWI with the TV prior yields the best results, with RMSE dropping below the CRLB bound especially at low photon counts.
This does not violate the CRLB, because the plotted bound is derived for an ideal unbiased estimator under the Poisson likelihood, whereas the TV-regularized PWI is a biased maximum a posteriori (MAP) estimator: it incorporates prior smoothness constraints, allowing its mean-squared error to fall below the unbiased bound.
Figure \ref{Fig2}C shows example reconstructions at different photon levels.
At the lowest tested level (1.58 mean photons), the PWI with the TV prior produces markedly better results than the SHWFS.
Additional details and phase resolution analysis are provided in the Supplementary Material.

To experimentally validate our method, we built the coded-detection setup. A polarized CW green laser (CPS532, Thorlabs) served as the light source, with neutral density filters used to attenuate the beam and create low photon conditions. The laser illuminated a USAF phase target (refractive index: 1.52, main feature height: 200 nm, Benchmark technologies), whose phase profile serves as the ground truth for our reconstruction. The phase modulation was performed by a spatial light modulator (SLM) (E19x12-400-800-HDM8, Meadowlark Optics), and corresponding modulated images were sequentially captured by an emCCD camera (PIMAX 4: 512EM, Teledyne Princeton Instruments). 
SLM phase diversity patterns were generated from low-resolution random matrices and interpolated to the full SLM resolution, producing smooth phase maps that reduce SLM crosstalk effects \cite{wu2019wish}. Refer to the photograph of the experimental setup provided in the Supplementary Material.

\begin{figure}[t]
\centering\includegraphics[width=0.60\textwidth]{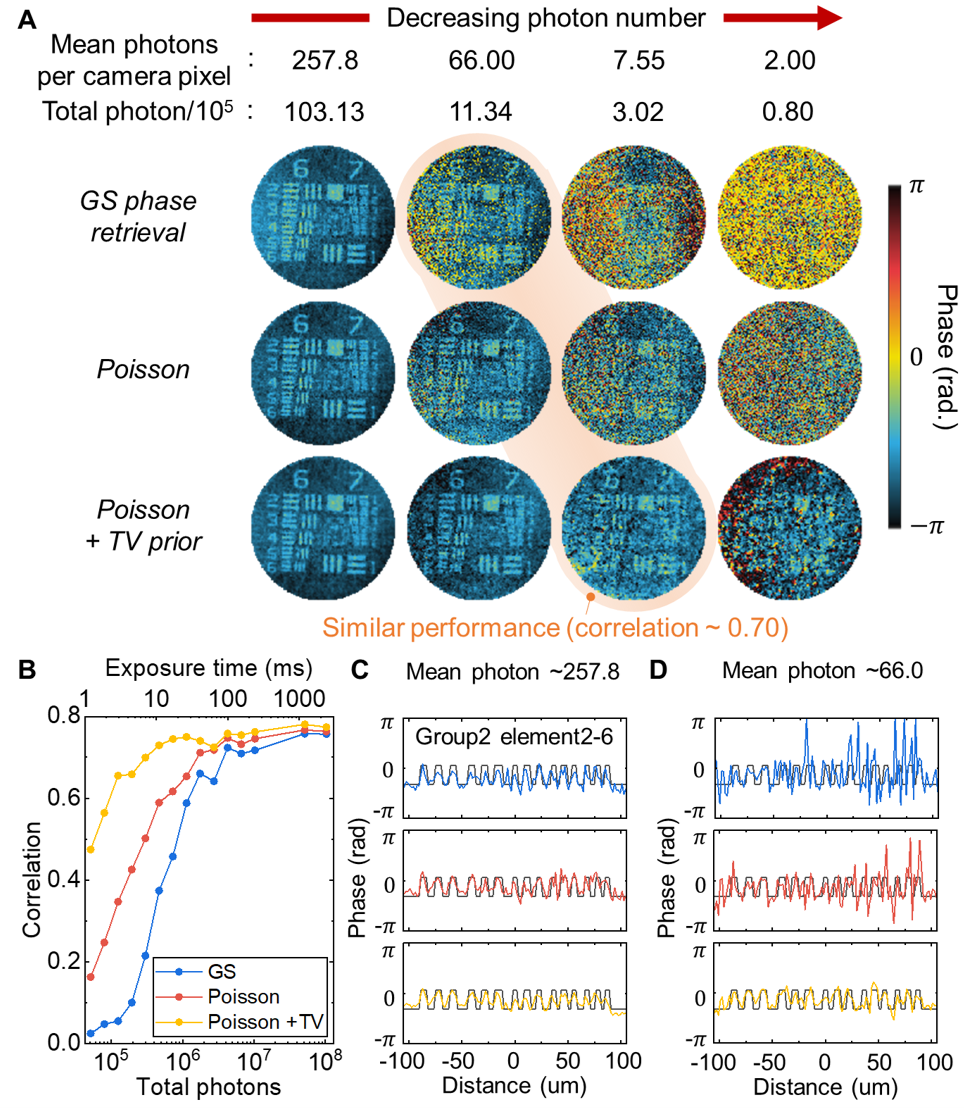}
\caption{
Comparative analysis of phase reconstruction across varied photon levels. A. Phase retrieval methods; GS, Poisson, and Poisson with TV prior are shown at total exposure times of 240, 26.4, 7.0 and 1.9 ms, correlating with cumulative mean
photons per pixel and total photons from 24 measurements. B. Correlation plots between ground truth and reconstructed phases demonstrate fidelity across different photon settings. C-D. Phase cross sections along elements 2–6 in Group 6 of the USAF resolution target, shown for mean photon levels of 257.8 (C) and 66.0 (D). Each panel displays, from top to bottom, GS, PWI, and PWI+TV results with ground truth profile (black).
Reconstructions used 24 SLM phase patterns and 400 iterations.
} 
\label{fig3}
\end{figure}
Figure \ref{fig3} compares phase reconstructions at various photon levels for three algorithms using the USAF phase target at the image plane, a configuration commonly used in microscopy and bio-imaging applications. Gaussian noise based GS algorithm, PWI with no prior, and PWI with the TV prior are compared, each using 24 SLM phase patterns.
Dark noise was removed from each input intensity profile by subtracting the average of 50–100 signal-free measurements. 
The cumulative photon counts across all measurements were determined using a photodiode (S120VC, Thorlabs), with detailed photon quantification provided in the Supplementary Material.
Figure \ref{fig3}A shows the qualitative performance gap between the methods across photon levels. The GS algorithm fails in photon-limited conditions ($<$7.55 mean photons per pixel), whereas the Poisson-based approaches maintain high reconstruction fidelity down to 2.0 photons per pixel, particularly when the TV prior is applied. Remarkably, while using only 7.55 photons per pixel, PWI with the TV prior achieves phase reconstruction quality comparable to GS at 66.0 photons per pixel---a reduction in photon budget by a factor of approximately 8.7, enabling substantially faster measurements.

Figure \ref{fig3}B shows the correlation between the reconstructed phases and the ground truth, computed as $\frac{|\Bar{g} \cdot x|}{||g||\cdot||x||}$, where $g$ is the ground truth and $\Bar{g}$ its complex conjugate \cite{seifert2023maximum}.
The Poisson-based method exhibits higher correlation than GS across the entire photon scanning range, with further improvements
when the TV prior is applied. Cross-sectional
analyses at mean photon numbers of 257.8 (Figure \ref{fig3}C)
and 66.0 (Figure \ref{fig3}D) further support these findings---while GS reconstructions are dominated by noise, especially at low photon levels (Figure \ref{fig3}D), Poisson-based result with the TV prior closely follow the true phase with reduced fluctuations. 
In particular, at 66.0 mean photons per pixel (Figure~\ref{fig3}D), the GS reconstruction can only reliably resolve up to Group~6, Element~2 of the USAF 1951 target, whereas PWI with TV resolves features up to Group~7, Element~1.
Using the standard USAF relation ($\text{resolution (lp/mm)} = 2^{\text{Group} + (\text{Element}-1)/6}$), this corresponds to an improvement from $71.8$ to approximately $128$ lp/mm, i.e., about a $1.8\times$ gain in the resolvable spatial frequency at the same photon budget.
These results demonstrate that incorporating the correct low-photon statistics and appropriate image priors enables robust phase reconstruction in photon-starved regimes, offering a pathway to faster and more efficient low-light imaging.

\begin{figure*}[t]
\centering\includegraphics[width=1.00\textwidth]{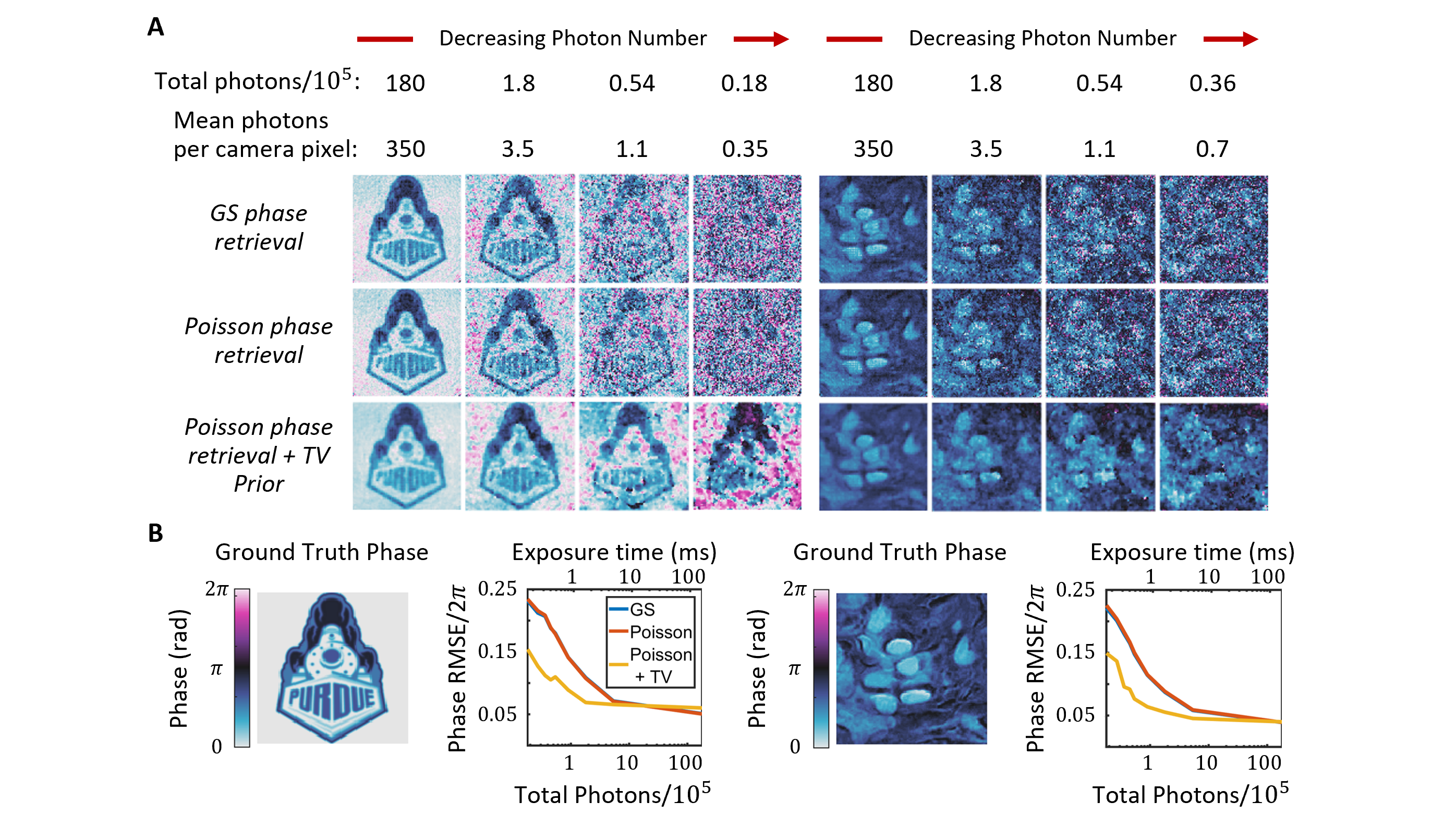} % Previous one: Fig4
\caption{
Phase retrieval tested across varying photon levels using a Purdue logo and stained tissue images displayed on the SLM. Phases are retrieved from 16 images with different SLM patterns. Total photon count from all images has an uncertainty of ±13\%. Each retrieval process involves 700 iterations. Ground truth phases and corresponding RMSE errors are shown in (B).
}
\label{fig4}
\end{figure*}

Next, we evaluate the phase RMSE using the SLM to generate a known phase target at the pupil plane, allowing tests with more complex structures. 
This setup reflects applications where pupil plane phase estimation is critical---for example, coherent diffraction imaging, where the target phase lies in the Fourier domain, or ground-based astronomical imaging, where turbulence-induced wavefront distortion originates at the system pupil.
The desired ground truth phase was superposed with each of the 16 SLM diversity patterns used for retrieval. 
Two targets were tested, a Purdue logo and stained biological tissue.
Figure \ref{fig4}A shows the retrieved phase for both targets. Unlike in Figure \ref{fig3}, the performance gap between GS and PWI without TV is negligible, with both methods yielding nearly identical performance across photon levels.
At an extremely low photon level of 0.35 photons/pixel, both GS and PWI without TV achieve 0.24 RMSE/2$\pi$, whereas adding the TV prior reduces the error to 0.15 RMSE/2$\pi$,---$1.6\times$ reduction---highlighting its consistent effectiveness in mitigating noise under photon-limited conditions.

We attribute the smaller performance gap between GS and PWI without TV in Figure \ref{fig4} to the target’s location in the optical system. In Figure \ref{fig3}, the target is at the image plane, where Poisson statistics can directly improve per-pixel intensity estimation and thereby aid phase recovery. In Figure \ref{fig4}, however, the target lies at the pupil plane, where the Fourier mapping spreads each image-plane pixel across the pupil, reducing the impact of localized Poisson-driven intensity estimation improvement.
This aligns with prior studies showing that the benefit of Poisson modeling depends strongly on the system matrix---being significant \cite{li2021poisson} in some settings, but minimal in lensless or pupil-plane configurations \cite{katkovnik2017phase}, which highlight its dependence on target position. Taken together, these two configurations clarify when Poisson modeling offers meaningful improvement: it provides substantial gains when the target lies at the image plane, but its influence naturally diminishes at the pupil plane, where the TV prior becomes the primary source of enhancement. Nevertheless, the combined Poisson–TV approach remains effective and robust across both regimes in the low-photon setting.

\section{Conclusion}
Several strategies could further enhance PWI. Using a Gaussian-Poisson mixed noise model and advanced filtering techniques such as BM3D can improve its accuracy \cite{seifert2023maximum, dabov2006image, katkovnik2017computational}. Deep learning can handle model imperfections and replace heuristic priors, provided a reliable dataset is available \cite{ma2024admm, cai2021data}.
For applications such as bio-imaging and semiconductor inspections, which involve single-wavelength setups, our method is readily used. To adapt for broadband light, however, it must handle partial coherence using spectral slicing \cite{bao2008phase, katkovnik2023admm}, for example.

In conclusion, we have successfully demonstrated phase retrieval with a few photons per camera pixel using a coded detection scheme. Our PWI method achieves wavefront estimation accuracy close to its theoretical limits and consistently outperforms the Gerchberg-Saxton algorithm in low photon experiments. Using a Poisson based loss function and TV regularization, we achieved up to 1.6$\times$ reduction in phase RMSE and 1.8$\times$ gain in effective phase resolution in low photon conditions.

\begin{backmatter}
\bmsection{Funding}
This work was supported by the Air Force Office of Scientific Research (AFOSR, grant FA9550-24-1-0345) and the Technology Innovation Program (IRIS no. RS-2024-00419426) funded by the Ministry of Trade, Industry and Energy (MOTIE, Korea).

\bmsection{Disclosures}
The authors declare no conflicts of interest.

\medskip

\bmsection{Data Availability}
Data underlying the results presented in this paper are
available from the corresponding authors upon reasonable request.

\bmsection{Supplemental document}
See Supplement 1 for supporting content.

\end{backmatter}

%%%%%%%%%%%%%%%%%%%%%%% References %%%%%%%%%%%%%%%%%%%%%%%%%

%Add references with BibTeX or manually 
%\cite{Zhang:14,OPTICA,FORSTER2007,Dean2006,testthesis,Yelin:03,Masajada:13,codeexample}.

%%%%%%%%%% If using BibTeX:
\bibliography{reference_Main}


\section*{Supplementary material}

# Poisson wavefront imaging in photon-starved scenarios: supplemental document

**Contents**

## 1. CODED-DETECTION SETUP

Our Poisson Wavefront Imaging (PWI) setup uses a coded detection approach with multiple phase diversity (Figure S1) [1, 2]. The system starts with a continuous wave (CW) green laser (CPS532, central wavelegth: 532 nm, average power: 4.5 mW, Thorlab). The light is horizontally polarized and then attenuated by a neutral density (ND) filter to mimic a low-photon environment. This beam then illuminates a USAF phase target (refractive index: 1.52, main feature height: 200 nm, Benchmark technologies). In the coded-detection setup, the light first passes through a 20× objective lens, a beam splitter (BS), and a spatial light modulator (E19x12-400-800-HDM8, 1920×1200 pixels with 8 um pixel pitch, Meadowlark Optics), where it undergoes phase modulation. The modulated light is then focused by a tube lens (Focal length = 25 cm) and captured by an EM-CCD camera (PIMAX 4: 512EM, 512×512 pixels with 23.6 um pixel pitch, Teledyne Princeton Instruments).

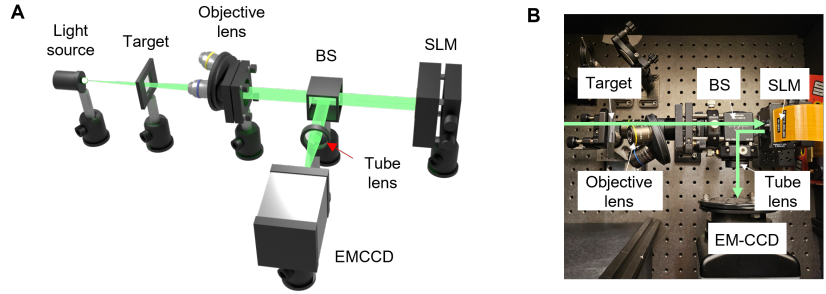

**Fig. S1.** PWI coded-detection setup. (A) Diagram and (B) Photograph of the experimental setup. This setup features a light source, target, objective lens, beam splitter (BS), spatial light modulator (SLM), tube lens and electron-multiplying charge-coupled device (EM-CCD).

## 2. ANGULAR SPECTRUM METHOD

We use an angular spectrum method (ASM), denoted as $\mathcal{P}_z$, to simulate the e-field propagation over a distance $z$. The ASM operator is defined as:

$$e_{\text{out}}(x,y) = \mathcal{P}_z e_{\text{in}}(x,y) = F^{-1}\left[H(f_x, f_y, z) \cdot F\left[e_{\text{in}}(x,y)\right]\right], \tag{S1}$$

where $e_{\text{in}}(x,y)$ is the input e-field, $e_{\text{out}}(x,y)$ is the propagated field, and $F$ and $F^{-1}$ represent the Fourier and inverse Fourier transform, respectively. The transfer function $H(f_x, f_y, z)$ for distance $z$ is:

$$H(f_x, f_y, z) = \begin{cases} \exp\left(2i\pi z/\lambda \sqrt{1 - \lambda^2(f_x^2 + f_y^2)}\right), & \text{if } f_x^2 + f_y^2 \leq 1/\lambda^2, \\ 0, & \text{otherwise.} \end{cases} \tag{S2}$$

Here, $f_x$ and $f_y$ are spatial frequency components, and $\lambda$ is the wavelength. Note that ASM performs pointwise multiplication between $H(f_x, f_y, z)$ and $F[e_{in}(x, y)]$ in the frequency domain. This means that after transforming efield $e_{in}(x, y)$ into the Fourier domain, the relationship with the transfer function $H(f_x, f_y, z)$ can be expressed through matrix multiplication. This is a useful property for implementing the update rules derived in the next chapter.

For simulating field propagation between optical elements, such as from the SLM to an imaging lens, and from the imaging lens to a detector, we use matrix form operation:

$$u_i = A_i a = \mathcal{P}_l \left\{ \mathcal{P}_l \left[ \Phi_i \cdot a \right] \cdot L \right\}, \tag{S3}$$

where $u_i$ is the resulting image plane e-field, $a$ is the pupil plane target e-field, $\Phi_i$ is the $i$-th SLM phase diversity pattern with aperture constraint, and $L$ is the tube lens phase with focal length $l$. Similarly, we can define wave propagation operator $A_0$ between target and SLM as follows:

$$a = A_0 x = \mathcal{P}_l \left\{ \mathcal{P}_l \left[ x \right] \cdot L \right\}, \tag{S4}$$

Here, $x$ represents the target plane e-field. We use the tube lens phase $L$ instead of an objective lens to maintain consistent magnification of the target e-field $x$ in our measurements.

## 3. UPDATE RULE DERIVATION

To aid in understanding this chapter, table S1 summarizes the key variables and parameters used in the analysis.

In low-light imaging, it is essential to use a statistical model that accurately reflects the sparse photon counts. A Gaussian approximation is often used when photon levels are high, as a Poisson distribution approaches Gaussian behavior for large mean values. However, in low photon conditions, a Gaussian model fails to accurately represent photon statistics, necessitating the use of a Poisson model. For a camera pixel with mean photon rate $\mu_{ik}$, the probability of measuring $n_{ik}$ photons is: $P(n_{ik}|\mu_{ik}) = \frac{\mu_{ik}^{n_{ik}} e^{-\mu_{ik}}}{n_{ik}!}$. Here, $i$ and $k$ are indices for coded-detection measurement and detector pixel, respectively. Using Bayes' theorem, the posterior probability for $\mu_{ik}$ given $n_{ik}$ can be expressed as: $P(\mu_{ik}|n_{ik}) = \frac{P(n_{ik}|\mu_{ik})P(\mu_{ik})}{P(n_{ik})}$. Assuming independence across pixels, the overall probability for the mean rates of all pixels is:

$$P(\mu|n) = \prod_{i,k} \frac{P(n_{ik}|\mu_{ik})P(\mu_{ik})}{P(n_{ik})}. \tag{S5}$$

To estimate the most probable scene from the measurements $n$, we seek the maximum a posteriori (MAP) estimate $\hat{\mu}$, which maximizes $P(\mu|n)$. This is equivalent to minimizing the negative log of $P(\mu|n)$:

$$\hat{\mu} = \underset{\mu}{\mathrm{argmin}} \left\{ \sum_{ik} -\log P(n_{ik}|\mu_{ik}) - \log P(\mu_{ik}) + \log P(n_{ik}) \right\} \tag{S6}$$

Since $\log P(n_{ik})$ is independent of $\mu$, it can be ignored in the optimization. The first term $-\log P(n_{ik}|\mu_{ik})$ becomes the main fidelity, while $-\log P(\mu_{ik})$ is replaced by regularization terms that incorporate prior knowledge, such as sparsity or smoothness.

In our coded detection phase retrieval approach, we employ specific phase modulations $\Phi_i$ to capture a series of images $I_i$. Let $g$ be the conversion factor between the camera output and photon counts, such that $n = gI$. This $g$ scaling is canceled in the optimization. We introduce $u_i$,

**Table S1.** Variables and parameters for phase reconstruction update rule.

| Symbol | Description | Symbol | Description |
| --- | --- | --- | --- |
| $P(n\|\mu)$ | Poisson probability of $n$ with mean $\mu$ | $A_i$ | Wave propagator from SLM (pupil) to detector plane |
| $n_{i,k}$ | $i$-th photon count at $k$-th detector pixel | $A_0$ | Wave propagator from target to SLM (pupil) plane |
| $\mu_{i,k}$ | Mean photon rate for $(i, k)$ measurement | $\Psi$ | TV image gradient operator |
| $g$ | Photon–intensity scaling factor | $T_x$ | Image gradient of $x$ |
| $I_{i,k}$ | Measured intensity at $(i, k)$ pixel | $\tau$ | Regularization weight for $T_x$ |
| $u_i$ | Complex field at detector plane | $\gamma_1$ | Penalty coefficient for $u_i - A_i a$ |
| $a$ | Complex field at SLM (pupil) plane | $\gamma_2$ | Penalty coefficient for $T_x - \Psi x$ |
| $x$ | Complex field at target plane | $\gamma_3$ | Penalty coefficient for $a - A_0 x$ |

the electric field at the detector plane, and assume $\mu_{ik} = |u_{ik}|^2$. Making these substitutions, the optimization problem becomes:

$$\{\hat{u}_i\} = \underset{\{u_i\}}{\operatorname{argmin}} \sum_{i=1}^{M} \sum_{k=1}^{K} \{-2I_{ik} \log(|u_{ik}|) + |u_{ik}|^2\} \tag{S7}$$

Here, the cost function depends only on the magnitude of the complex field, reflecting that intensity measurements contain no phase information. To include the phase, we consider the fields at the SLM plane ($a$) and the target plane ($x$). These fields are related through propagation equations: $a = A_0 x$ represents propagation from the target to the pupil plane, and $u_i = A_i a$ includes the $i$-th SLM phase $\Phi_i$ and propagation to the sensor plane. These conditions act as constraints for this optimization problem. And we integrate a total variation (TV) prior as regularization, encouraging smoothness in the target efield $x$, which gives the following constraint problem:

$$\underset{\{u,a,x,T_x\}}{\operatorname{argmin}} \sum_{i=1}^{M} \sum_{k=1}^{K} \{-2I_{ik} \log(|u_{ik}|) + |u_{ik}|^2\} + \tau ||T_x||_1 \tag{S8}$$
$$\text{s.t.} \quad T_x = \Psi x, \, u_i = A_i a, \, a = A_0 x, \, i = 1, ..., M.$$

Here, $\Psi$ is a differential operator for the TV prior, $\tau$ is a balancing coefficient for the TV prior, $T_x$ is an auxiliary variable subject to TV regularization, and $||\cdot||_1$ is the L1 norm. Note that the TV prior is applied directly to the complex field $x$. We adopt this formulation because complex-field TV avoids phase unwrapping, stabilizes both amplitude and phase under low-photon noise, and admits a closed-form proximal update within the ADMM framework that follows.

We convert the constrained optimization to an unconstrained one using augmented Lagrangian methods. The unconstrained loss function is then:

$$\begin{aligned} L = \sum_{i=1}^{M} \sum_{k=1}^{K} &\{-2I_{ik} \log(|u_{ik}|) + |u_{ik}|^2 + \frac{\gamma_1}{2} ||A_i a - u_i + \lambda_{1i}||^2\} \\ &+ \tau ||T_x||_1 + \frac{\gamma_2}{2} ||\Psi x - T_x + \lambda_2||^2 \\ &+ \frac{\gamma_3}{2} ||A_0 x - a + \lambda_3||^2 \end{aligned} \tag{S9}$$

The penalty coefficients $\gamma_1, \gamma_2, \gamma_3$ enforce adherence to the constraints, while $\lambda_1, \lambda_2, \lambda_3$ are the corresponding dual variables (Lagrange multipliers) introduced in the augmented Lagrangian formulation to facilitate convergence and maintain constraint consistency. $\gamma_1$ depends on the signal level, where $\gamma_1 \to 0$ represents a noiseless situation, and $\gamma_1 \to \infty$ represents a noise-only scenario. To solve for $\{u_i\}_{i=1,...,M}$, $a$, $x$, and $T_x$, we apply alternating minimization approach, which optimizes each variable sequentially while holding the others fixed. This iterative process updates each variable at the $j$-th iteration as follows:

$$\{u_i^{(j+1)}\} = \underset{\{u_i\}}{\operatorname{argmin}} L(\{u_i\}, a^{(j)}, x^{(j)}, T_x^{(j)}) \tag{S10}$$

$$a^{(j+1)} = \underset{a}{\operatorname{argmin}} L(\{u_i^{(j+1)}\}, a, x^{(j)}, T_x^{(j)}) \tag{S11}$$

$$x^{(j+1)} = \underset{x}{\operatorname{argmin}} L(\{u_i^{(j+1)}\}, a^{(j+1)}, x, T_x^{(j)}) \tag{S12}$$

$$T_x^{(j+1)} = \underset{u_i}{\operatorname{argmin}} L(\{u_i^{(j+1)}\}, a^{(j+1)}, x^{(j+1)}, T_x) \tag{S13}$$

Instead of using methods like stochastic gradient descent to solve the problem, we use closed-form solutions at stationary points to accelerate convergence [3, 4]. For analytical gradient calculations, each operation must be expressed pointwise (or as diagonal matrix multiplications) in a specific basis. As we discussed in previous chapter, the ASM operator express a pointwise multiplication in Fourier domain, allowing us to represent the loss function entirely in matrix form. This yields closed-form update rule through stationary point, as detailed in Table 1 of the main text. On the other hand, dual variables are updated using gradient ascent.

In our analysis, the ADMM penalty parameters were coupled as:

$$\gamma_2 = 5 \times 10^{-1}\gamma_1, \quad \gamma_3 = 5 \times 10^{-1}\gamma_1, \quad \tau = 5 \times 10^{-2}\gamma_2$$

and only $\gamma_1$ was adapted to the photon level, ranging from $10^{-4}$ to $10^{-2}$. These values were chosen heuristically after linearly normalizing each raw intensity frame so that its pixel values lie in the range $[0,1]$. Intuitively, $\gamma_1$ controls the trade-off between trusting the noisy measurements and the propagated prior estimate: smaller $\gamma_1$ gives more weight to the measurements, which is preferable when the data are less noisy, whereas larger $\gamma_1$ places more weight on the previous estimate and suppresses noise in strongly photon-limited data.

For the stopping criterion, we empirically determined the required number of iterations in the large-photon regime ($10^4$ mean photons per detector pixel), selecting the point at which the reconstruction approached convergence in simulation (approximately 0.01 RMSE/$2\pi$). Following this procedure, we used about 400 iterations when the target was in the image plane (Figure 3 in the main text) and about 700 iterations when the target was in the pupil plane (Figure 4 in the main text).

## 4. CONVERGENCE COMPARISON

In this section, we compare the deterministic, closed-form update rules of the proposed PWI solver with a stochastic gradient–based optimization using Adam. Unlike PWI, which sequentially solves each subproblem in closed form and therefore guarantees a monotonic decrease of the augmented objective, Adam updates rely solely on local gradient information (even with its momentum) and are susceptible to stagnation in poor local minima.

To assess reconstruction quality in this test, we employ the Blind/Referenceless Image Spatial Quality Evaluator (BRISQUE), a widely used no-reference image quality metric based on natural scene statistics. Lower BRISQUE values indicate higher perceptual quality. As shown in Fig. S2, Adam achieves a rapid initial decrease in BRISQUE score but quickly saturates at around 200 iterations, reflecting its inability to escape early local minima. In contrast, the PWI solver continues to improve throughout the iterations, yielding significantly cleaner amplitude and phase reconstructions.

These results highlight the practical advantage of PWI's deterministic update structure: by enforcing physical propagation consistency, Poisson denoising, and TV regularization through alternating closed-form updates, PWI achieves substantially higher reconstruction fidelity compared with stochastic gradient descent.

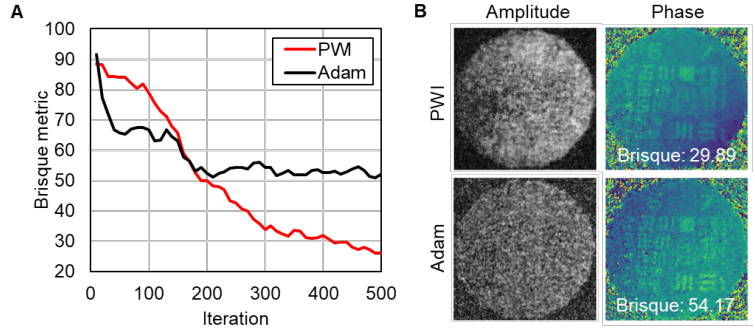

**Fig. S2.** Convergence comparison between PWI and Adam. (A) BRISQUE score of reconstructed phase versus iteration for the PWI and Adam solver (learning rate = 0.01). Lower values indicate higher perceived image quality. While Adam rapidly reaches a plateau around 200 iterations, PWI continues to improve throughout the optimization. (B) Amplitude and phase reconstructions after 500 iterations. PWI yields more natural amplitude profiles and cleaner phase reconstructions, whereas Adam suffers from high-frequency phase noise and residual artifacts. Note that both approaches optimize the same loss function, including the TV-regularization term.

In addition to its improved reconstruction accuracy, PWI also provides superior computational efficiency compared with stochastic gradient–based optimization. The computational bottleneck

of both methods is the ASM operator, which requires two FFTs per propagation. For an image grid of size $K$, the total computational complexity of PWI scales as $O(IMK \log K)$, which is linear in both the number of iterations $I$ and the number of SLM patterns $M$. In contrast, stochastic gradient methods introduce additional overhead to track gradient statistics—particularly when using momentum-based optimizers such as Adam—resulting in a significantly higher per-iteration cost. Figures S3A and S3B experimentally verify these trends where both PWI and Adam show linear scaling behavior with respect to SLM patterns $M$ and total iteration $I$, while PWI present almost $1.7\times$ speed improvement.

Note that this speed improvement is achieved while simultaneously delivering substantially more accurate amplitude and phase reconstructions, underscoring the practical advantage of PWI's deterministic closed-form updates over stochastic gradient descent.

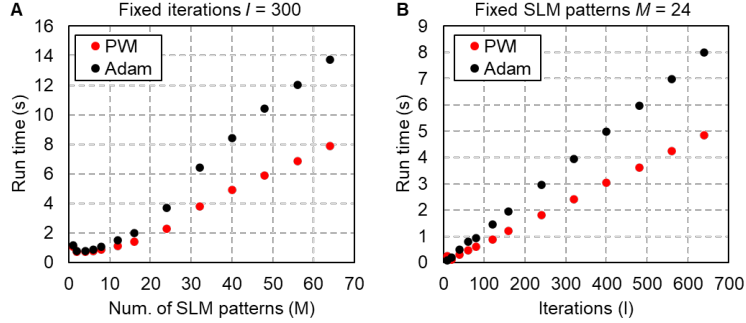

**Fig. S3.** Running time comparison between PWI and Adam. (A) running time versus SLM patterns number $M$ for PWI and Adam (300 iterations, image grid size of $K = 400 \times 400$). (B) running time versus the number of iterations $I$ for fixed $M = 24$. Across both experiments, the proposed PWI solver consistently achieves lower runtime than Adam solver.

## 5. FISHER INFORMATION ANALYSIS

Fisher information quantifies the amount of information a measurement system provides about unknown parameters. We compare the phase information content of PWI with flat phase diversity, random phase diversity, and Shack–Hartmann wavefront sensing (SHWFS). By calculating the Cramer–Rao lower bound (CRLB), we establish the theoretical precision limits of these systems. Two performance metrics are considered: (i) phase estimation error and (ii) phase image resolution.

### A. Scenario 1: Theoretical bound on wavefront estimation

#### A.1. Theory

We evaluate how accurately the system detects pixel-wise phase values. The test scene is a 40x40 complex efield. Its phase, **x**, forms a "Purdue P" letter, while the amplitude is flat. The parameters of interest are the pixelated phase values $\{\theta_p\}_{p=1,...,P}$, with $P = 1600$. To quantify performance, we compute the single-photon Fisher information matrix (FIM) $J$. Each entry $J_{pq}$ captures the information about phase pixels $p$ and $q$. The FIM is obtained in the detector plane by summing contributions over all measurement pixels $k$ and all coded detection patterns $i$:

$$J_{pq} = \sum_{k=1}^{K} \sum_{i=1}^{M} \frac{1}{\mu_{ik}} \frac{\partial \mu_{ik}}{\partial \theta_p} \cdot \frac{\partial \mu_{ik}}{\partial \theta_q}, \tag{S14}$$

Where $K$ is the number of detector pixels, $M$ is the number of coded-detection frames, $\mu_{ik}$ represents the single-photon distribution at detector pixel $k$ for the $i$-th coded-detection measurements (i.e. $\sum_k \sum_i \mu_{ik} = 1$). The CRLB matrix, $\sigma^2$, iis defined as the inverse of the Fisher information matrix (FIM) after removing one row and column to eliminate the piston ambiguity. Each diagonal element, $\sigma_{pp}$ gives the theoretical minimum error for phase pixel $p$. For a total of $N$ photons, the average standard deviation is $\sigma_{min}(N) = \sqrt{\frac{1}{N(P-1)} \sum_{p=1}^{P-1} \sigma_{pp}^2}$, which quantifies the minimum achievable uncertainty in estimating the pixelated phase values.

5

### A.2. Simulation model

- PWI: Objective lens ($f = 5$ cm) collimates **x**, SLM imposes phase modulation, and a tube lens ($f = 5$ cm) focuses onto a 14 $\mu$m-pixel camera. The $40 \times 40$ field is zero-padded to $200 \times 200$.

- SHWFS: Microlens array with 14 $\mu$m pitch and 80 $\mu$m focal length. Each lenslet forms an $11 \times 11$ PSF grid on the camera, yielding $440 \times 440$ intensity data. Phase gradients are extracted via centroid shifts and integrated using the Southwell algorithm [5], following the update equation:

$$\phi_{n,m}^{t+1} = \sum_{j=-1}^{1} \sum_{i=-1}^{1} I_{n+i,m+j} \cdot [\phi_{n+i,m+j}^{t} + \frac{\Delta}{2} \cdot \{\frac{\partial \phi}{\partial x_{n,m}} + \frac{\partial \phi}{\partial x_{n+i,m+j}}\}] \tag{S15}$$

Where, $\phi_{n,m}^{t}$ is the wavefront retrieved at grid $(n, m)$ during iteration $t$, $\Delta$ is grid spacing, $I_{n+i,m+j}$ is the measured intensity in the neighboring region, and $\frac{\partial \phi}{\partial x_{n,m}}$ denotes the measured phase gradient at $(n, m)$. For stable convergence, the number of iterations is chosen larger than the squared number of lenslets.

## B. Scenario 2: Phase image resolution

### B.1. Theory

Classical definitions of resolution—such as the diffraction limit or Rayleigh criterion—are rooted in the intensity domain and assume deterministic conditions. These metrics fail to capture resolution degradation under photon-starved conditions, where statistical noise dominates. In this scenario, we redefine resolution through Fisher information in the phase domain, making resolution explicitly dependent on the photon number. This framework allows us to quantify how photon-limited measurements constrain resolvability and to directly compare the performance of coded detection against flat phase diversity. We model the target as two phase point sources separated by distance $\theta$:

$$\mathbf{x}(x, y, \theta) = A(x, y) \exp[j\{\mathbf{1}_A(x, y) + \mathbf{1}_A(x + \theta, y)\}] \tag{S16}$$

Where $A(x, y)$ is normalized single-photon amplitude, and $\mathbf{1}_A(x, y)$ is an indicator function for the source locations:

$$\mathbf{1}_A(x, y) = \begin{cases} 1 & \text{if } x = 0 \text{ and } y = 0, \\ 0 & \text{otherwise.} \end{cases} \tag{S17}$$

The parameter of interest is the separation $\theta$. Using Eq. S14, we compute the Fisher information $J(\theta)$ for a single photon. The CRLB gives the minimum standard deviation of any unbiased estimator: $\sigma(\theta, N) = \frac{1}{\sqrt{NJ(\theta)}}$ where $N$ is the photon count. This bound establishes the achievable resolution limit and makes explicit the joint dependence on $\theta$ and $N$.

To further interpret this definition, we evaluate the resolving power of the system by plotting the Fisher information $J(\theta, N)$ and the associated standard deviation $\sigma(\theta, N)$ over a range of separations $\theta$ and photon counts $N$ (Figs. S4A and B). The 2D maps show that smaller separations or lower photon numbers reduce $J(\theta, N)$ and increase $\sigma(\theta, N)$, indicating diminished resolution capacity. Resolution is achieved when $\sigma(\theta, N) \leq \theta$, meaning the estimation bound is tighter than the true separation. The minimum resolvable distance $\Delta(N)$ is thus defined as the separation $\theta$ that satisfies $\sigma(\theta, N) = \theta$, establishing a complementary dependence between photon number $N$ and achievable resolution. Figure S4C summarizes $\Delta(N)$ for different systems, showing that coded detection with phase diversity outperforms direct imaging with flat phase diversity. This confirms that coded detection not only improves phase estimation but also enhances phase image resolution under photon-starved conditions.

### B.2. Simulation model

- PWI: Objective lens ($f = 20$ cm) collimates **x**, SLM imposes phase modulation, and a tube lens ($f = 20$ cm) focuses onto a 16 $\mu$m-pixel camera. Random phase diversity patterns are applied at the SLM, with pitch sizes set to 5 and 10 pixels.

- DI: The same optical setup is used, but no phase diversity is introduced at the SLM, corresponding to direct imaging without coded modulation.

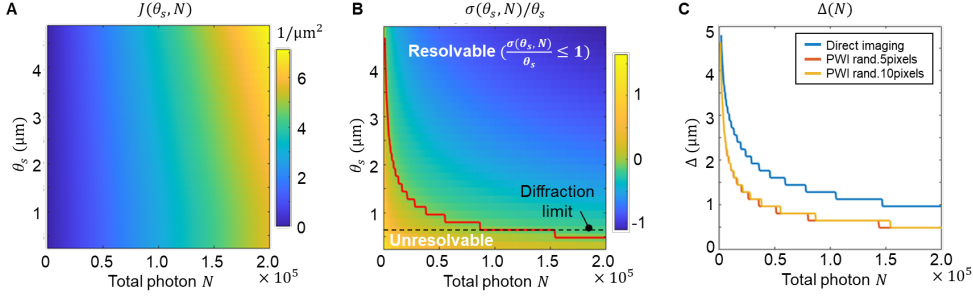

**Fig. S4.** Theoretical limit of image resolution vs. total photon for direct imaging and coded-detection architecture. A-B. Fisher information and CRLB with respect to total photon and separation level $\theta$ for coded-detection system. C. Image resolution comparison.

## 6. PHOTON NUMBER MEASUREMENT

To determine the number of photons in each measured image, we first measured the incident power without neutral density (ND) filters using a silicon photodiode (S120VC, Thorlabs). The photodiode was placed directly in front of the camera to ensure that the entire beam was collected on its active area. It was also used to calibrate the ND filters, yielding a total attenuation factor of $1.8 \times 10^4 \pm 12\%$ in our setup.

The total number of detected photons under a given condition is then

$$N_{\text{phot}} = \frac{P t_{\text{exp}}}{A_{\text{ND}}} \frac{\lambda}{hc} \frac{1}{\eta}, \tag{S18}$$

where $P$ is the measured incident power (no ND filter), $A_{\text{ND}}$ the attenuation factor of the ND filters, $t_{\text{exp}}$ the exposure time, $\eta$ the detector quantum efficiency, and $\lambda/(hc)$ the photon energy conversion factor. Finally, the mean photon count per pixel is obtained by dividing $N_{\text{phot}}$ by the number of illuminated pixels when no phase pattern is applied to the SLM.

## 7. PHASE RMSE CALCULATION

The standard definition of RMSE does not properly account for phase wrapping and may give inconsistent errors under global phase shifts. Such behavior is undesirable, since a global phase shift has no physical meaning. For example, if the true phase is 0 and the retrieved phase is $2\pi - 0.1$, the raw error appears large, although both phases differ by only 0.1 after a global shift.

To address this, we define the absolute phase error as

$$err_{\text{abs}}(\phi_1, \phi_2) = \begin{cases} |\phi_1 - \phi_2| & \text{if } |\phi_1 - \phi_2| \leq \pi, \\ 2\pi - |\phi_1 - \phi_2| & \text{if } \pi < |\phi_1 - \phi_2| < 2\pi, \end{cases} \tag{S19}$$

ensuring that the error never exceeds $\pi$, the maximum possible difference between two wrapped phases.

The phase RMSE between two phase images is then defined as

$$RMSE_{\text{phase}}(\phi_1, \phi_2) = \sqrt{\frac{1}{K} \sum_{k=1}^{K} err_{\text{abs}}(\phi_1(k), \phi_2(k))^2}, \tag{S20}$$

where $K$ is the number of pixels. In practice, we compute the phase RMSE for different relative global shifts between the ground truth and the retrieved phase. The minimum value across these shifts is reported as the final phase RMSE in the main text.

## 8. PATTERN NUMBER OPTIMIZATION

In the noiseless case, increasing the number of phase patterns and corresponding measurements can only improve or maintain performance. However, in realistic scenarios with noise, the advantage of more patterns must be balanced against reduced signal per image and additional

read noise. Each frame readout contributes read noise, so splitting the total signal into more frames can increase the effective noise level.

We experimentally tested the optimal number of SLM patterns under varying photon budgets, using a setup similar to Fig. 4 in the main text. The total photon number was kept constant across different pattern counts; therefore, using more patterns reduces the number of photons per image.

The results, shown in Fig. S5, indicate that the optimal number of patterns depends on the photon level. At low photon counts, fewer patterns are preferable to avoid weak individual measurements. At high photon counts, noise is less limiting, and using more patterns improves retrieval performance.

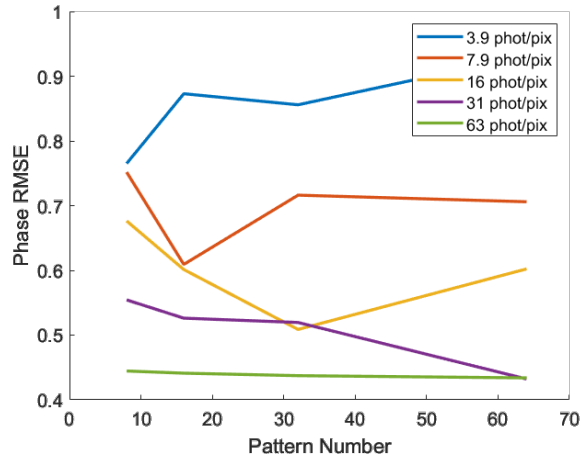

**Fig. S5.** Optimal pattern number analysis. The experimental setup matches Fig. 4 in the main text with the Purdue logo phase target. The total number of photons is fixed across all cases, so larger pattern numbers correspond to fewer photons per image. Here, the photon count per pixel is calculated from the retrieved phase target pixels.



\end{document}